\documentclass[fleqn,usenatbib]{mnras}
\hypersetup{linkcolor=DodgerBlue,citecolor=MediumBlue,filecolor=red,urlcolor=FireBrick}

\usepackage[T1]{fontenc}
\usepackage{graphicx}	
\usepackage{amsmath}	
\usepackage{natbib}
\usepackage{txfonts}
\usepackage{lscape}     
\usepackage{placeins}   
\usepackage{amssymb}	
\usepackage{multirow}
\usepackage{siunitx}
\usepackage{threeparttable}
\usepackage{cleveref}
\usepackage{pdflscape}
\usepackage{caption}
\usepackage{subcaption}    
\usepackage{float}
\usepackage{booktabs}
\usepackage{mathrsfs}
\usepackage{euscript}
\usepackage[svgnames]{xcolor}
\usepackage{orcidlink}
\usepackage{cleveref}

\DeclareRobustCommand{\VAN}[3]{#2}
\let\VANthebibliography\thebibliography
\def\thebibliography{\DeclareRobustCommand{\VAN}[3]{##3}\VANthebibliography}

\DeclareSIUnit \parsec 	    {pc}
\DeclareSIUnit \mparsec 	{mpc}
\DeclareSIUnit \eV 			{eV}
\DeclareSIUnit \keV 		{keV}
\DeclareSIUnit \Msun 		{M_\odot}
\DeclareSIUnit \year        {yr}

\newcommand{\MBH}{M_{{\rm BH}}} 
\newcommand{\Msun}{M_\odot} 
 
\title[Dynamics around a fermionic dark matter core]{The dynamics of S-stars and G-sources orbiting a supermassive compact object made of fermionic dark matter}

\author[V. Crespi et al.]{
V. Crespi,$^{1,2}$ \thanks{e-mail: valentinacrespi@fcaglp.unlp.edu.ar} \orcidlink{0009-0005-8190-3598}
C.~R.~Arg\"uelles,$^{1,2,3,4}$ \orcidlink{0000-0002-5862-8840}
E.~A.~Becerra-Vergara$^{3,4,5}$ \orcidlink{0000-0003-4848-1483}
M.~F.~Mestre$^{1,2}$ \orcidlink{0009-0001-9329-5260}
F. Pei\ss ker$^{6}$
\newauthor
J.~A.~Rueda$^{3,4,7,8,9}$ \orcidlink{0000-0003-4904-0014}
R. Ruffini $^{3,9,10}$
\\
$^{1}$ Instituto de Astrof{\'i}sica de La Plata (CONICET-UNLP), Paseo del Bosque S/N, La Plata (1900), Buenos Aires, Argentina\\
$^{2}$ Fac. de Ciencias Astron. y Geof\'isicas, Universidad Nacional de La Plata, Paseo del Bosque, B1900FWA La Plata, Argentina\\
$^{3}$ ICRANet, Piazza della Repubblica 10, I-65122 Pescara, Italy \\
$^{4}$ ICRA, Dip. di Fisica, Sapienza Universit\`a di Roma, P.le Aldo Moro 5, I-00185 Rome, Italy \\
$^{5}$ GIRG, Escuela de F\'isica, Universidad Industrial de Santander A. A. 678, Bucaramanga 680002, Colombia \\
$^{6}$ I. Physikalisches Institut der Universit\"at zu K\"oln, Z\"ulpicher Str. 77, D-50937 K\"oln, Germany \\
$^{7}$ ICRANet-Ferrara, Dip. di Fisica e Scienze della Terra, Universit\`a degli Studi di Ferrara, Via Saragat 1, I-44122 Ferrara, Italy \\
$^{8}$ Dip. di Fisica e Scienze della Terra, Universit\`a degli Studi di Ferrara, Via Saragat 1, I-44122 Ferrara, Italy \\
$^{9}$ INAF, Istituto de Astrofisica e Planetologia Spaziali, Via Fosso del Cavaliere 100, I-00133 Rome, Italy \\
$^{10}$ INAF, Viale del Parco Mellini 84, I-00136 Rome, Italy
}

\date{Accepted XXX. Received YYY; in original form ZZZ}
\pubyear{\the\year{}}

\begin{document}
\label{firstpage}
\pagerange{\pageref{firstpage}--\pageref{lastpage}}
\maketitle

\begin{abstract}
    Surrounding Sgr A*, a cluster of young and massive stars coexist with a population of dust-enshrouded objects, whose astrometric data can be used to scrutinize the nature of Sgr A*. An alternative to the black hole (BH) scenario has been recently proposed in terms of a supermassive compact object composed of self-gravitating fermionic dark matter (DM). Such horizon-less configurations can reproduce the relativistic effects measured for S2 orbit, while being part of a single continuous configuration whose extended halo reproduces the latest GAIA-DR3 rotation curve. In this work, we statistically compare different fermionic DM configurations aimed to fit the astrometric data of S2, and five G-sources, and compare with the BH potential when appropriate. We sample the parameter spaces via Markov Chain Monte Carlo statistics and perform a quantitative comparison estimating Bayes factors for models that share the same likelihood function. We extend previous results of the S2 and G2 orbital fits for $56$ keV fermions (low core-compactness) and show the results for $300$ keV fermions (high core-compactness). For the selected S2 dataset, the former model is slightly favoured over the latter. However, more precise S2 datasets, as obtained by the GRAVITY instrument, remain to be analysed in light of the fermionic models. For the G-objects, no conclusive preference emerges between models.  For all stellar objects tested, the BH and fermionic models predict orbital parameters that differ by less than 1\%. More accurate data, particularly from stars closer to Sgr A*, is necessary to statistically distinguish between the models considered.
\end{abstract}



\section{Introduction}

Near-infrared wavelength observations bypass the gas and dust components of the central pc of the Galaxy, revealing information about the stars orbiting the supermassive compact object, SgrA*.
Back to the early 1980s, a BH has been the main candidate to be evaluated by ESO in the Galactic Centre (see \cite{1987ARA&A..25..377G} for a review). This hypothesis was possible mainly thanks to the mid-infrared spectroscopy of ionized gas near Sgr A* at La Silla, which revealed velocity dispersions consistent with a central $\sim 3\times10^6 M_\odot$ compact object. Continuous efforts in the 1990s, \citet{1997MNRAS.284..576E}, and shortly after  \citet{1998ApJ...509..678G}, were able to measure independently different stars orbiting Sgr A* at mpc scales and named them as S/S0 stars. Thereafter, more than three decades of continuous collection of astrometric and spectroscopic data was possible thanks to the use of different instruments from different research teams and institutions \citep{2002Natur.419..694S,Rousset_2003,2005ApJ...620..744G,2008ApJ...689.1044G,Eisenhauer_2005,2009ApJ...692.1075G,2017ApJ...837...30G,2017ApJ...845...22P,2018A&A...615L..15G,2019Sci...365..664D}. 
These data allowed for a progressive constraint in the mass and distance to Sgr A*, inferred to be approximately $R_{\odot} = 8.277$ kpc and $M_{\rm BH} = \SI{4.297E6}{\Msun}$ \citep{2022A&A...657L..12G}. 

So far, the orbits of the S-cluster stars are the most reliable tracers of the gravitational field produced by Sgr A*. The astrometry data of the S-star orbits have been traditionally explained according to the supermassive black hole (BH) paradigm \citep{2005ApJ...620..744G, 2008ApJ...689.1044G, 2009ApJ...692.1075G, 2010RvMP...82.3121G, 2017ApJ...837...30G, 2018A&A...618L..10G, 2019Sci...365..664D,2020A&A...636L...5G}. 
Regardless, alternative paradigms falling either within or beyond the theory of General Relativity (GR) have been proposed for Sgr A* over the years 
(see, e.g., \citealp{2024A&ARv..32....3G}, for a recent review on the Galactic Centre).
Within the classical theory of GR, the most studied scenarios for Sgr A* include the case of self-gravitating systems made either of bosons such as boson stars \citep{2000PhRvD..62j4012T,2006PhRvD..73b1501G,vincent2016,vincent2021,2020MNRAS.497..521O,rosa2023imaging}, or made of dark matter fermions \citep{2002PrPNP..48..291B,2015MNRAS.451..622R,2016JCAP...04..038A,2016MNRAS.461.4295S,2018PDU....21...82A,2020A&A...641A..34B,2021MNRAS.505L..64B,2022MNRAS.511L..35A,2024MNRAS.534.1217P,2025arXiv250310870K}. 
The fermionic case, when worked out within the Ruffini-Arg\"uelles-Rueda (RAR) model \citep{2015MNRAS.451..622R}, or its more accurate extension allowing for particle evaporation \citep{2018PDU....21...82A}, represents an appealing scenario. The extended RAR model contains equilibrium solutions for the DM density profiles with a \textit{dense core}--\textit{diluted halo} morphology, where the dense fermion-core can mimic the central BH in SgrA* and the outer halo explains the Galaxy rotation curve \citep{2020A&A...641A..34B,2021MNRAS.505L..64B,2022MNRAS.511L..35A,2023Univ....9..372A,2025arXiv250310870K}. In the fermion-core alternative, compact enough configurations that explain the dynamics of the S-cluster stars, including its relativistic effects, i.e., gravitational redshift and periapsis precession \citep{2020A&A...641A..34B,2021MNRAS.505L..64B,2022MNRAS.511L..35A}, can also produce shadow-like features with sizes compatible with the measurements of the EHT collaboration when applied to Milky Way-like galaxies (\citealp{2024MNRAS.534.1217P}; see also \citealp{rosa2023imaging}, for the case of bosons).

The young S-star cluster is located in the innermost $\sim 40$ mpc, and its formation history is still a matter of debate \citep{2003ApJ...586L.127G,Yusef-Zadeh_2013,2014MNRAS.444.1205J,2020ApJ...896..100A}.
Among the features of the S-stars are their high velocities up to several thousands of km s$^{-1}$, with high eccentric orbits in some cases. 
Due to its brightness and data quality, the S2 star is the most relevant of the cluster. It has a relatively short orbital period of $P \sim 16$ yr around the Galactic Centre in a highly eccentric orbit $e \sim 0.88$, and is one of the brightest of the cluster $m_{\rm K} \sim 14$ \citep{2018A&A...615L..15G,2019Sci...365..664D}. 
It gained particular attention in 2018, when it had its second recorded passage through the pericentre, allowing for the measurement of different first-order GR effects. They include the gravitational Redshift, along with the transverse Doppler effect of special relativity \citep{2018A&A...615L..15G,2019Sci...365..664D} and soon after, the first measurement of the relativistic Schwarzschild precession of about $\sim 12$ arcmin per revolution \citep{2020A&A...636L...5G}.
 
Together with the tight S-cluster of young and bright stars around Sgr A*, exists a population of dust-enshrouded objects, known as G-sources, orbiting in the inner $0.1$ pc of the Galaxy \citep{2017ApJ...847...80W,2020Natur.577..337C}. On observational grounds, these object belong however to a different family of stars as compared with the S-stars: their astrometric data is not quantitative nor accurate enough to report any relativistic effect as in the case of the S-2 star. 
A somewhat controversial object is G2, detected in the last decade by \citet{2012Natur.481...51G}, where the authors interpreted it to be a gas cloud orbiting in a highly eccentric orbit towards Sgr A*, and expected to be tidally disrupted during its pericentre passage \citep{2012Natur.481...51G,Burkert_2012}. 
However, the object survived its closest approach to Sgr A*, giving rise to some dispute about its nature. 
In \citet{2019ApJ...871..126G}, the authors state that the core-less gas cloud G2 may be a knot in a larger gas stream, wherein the head, there is another cloud -- first identified by \citet{2005sao..conf..286C} -- called G1, that follows a similar orbit to that of G2. 
To explain the observed radial velocity of the last post pericentre data of G2 under the gas cloud hypothesis, \citet{2019ApJ...871..126G} assumed a drag force by an accretion flow since the solely gravitational potential of a central BH cannot explain the data.
However, other authors stated that the observed characteristics of G2 are more consistent with a stellar source surrounded by a gaseous-dusty envelope \citep{2013ApJ...768..108S,2014ApJ...796L...8W,Prodan_2015,2020A&A...634A..35P}, and that it is an entirely independent object from G1 \citep{2017ApJ...847...80W}. 
Indeed, some consensus in the literature has been reached on the later hypothesis after the detailed analysis on G2 made by \citet{2021ApJ...923...69P}, who pointed out that data smoothing (as the one used in \citealp{2019ApJ...871..126G}) significantly affects the kinematics analysis of DSO-like sources (Dusty S-cluster Objects).
They presented a new Keplerian-like orbit for the compact stellar source and two more individual sources in the claimed tail, with no need to impose a drag force on G2 as previously stated. Therein, it was argued that the ionized gas associated with G2 is located within the S-cluster and, therefore, can be disentangled from its emission.
Recently, \citet{2020Natur.577..337C} proposed four other sources that share the characteristics of G1 and G2 and named them G3, G4, G5 and G6. As defined by the authors, G objects present the properties of gas and dust clouds but follow the dynamics of stellar-mass objects. More specifically, there is a distinct source of Brackett-$\gamma$ emission; the emission region is spatially compact, with a relatively weak K-band continuum emission with very red K-L colors (indicating they are probably enshrouded by dust), and large proper motion and radial velocity shifts over time.\\
\indent
In this work, we aim to constrain the gravitational potential of Sgr A* when modeled by the fermionic \textit{core-halo} distribution. For this task, we will use astrometric data of S star and G objects, and compare the results with the BH hypothesis. We first performed a Markov Chain Monte Carlo (MCMC) sampling test with the available S2 orbit data. In this way, we extend the previous results of \citet{2020A&A...641A..34B}, exploring different (i.e., $mc^2=300$ keV) fermionic particle mass range, implying more compact fermion-cores relatively close to the gravitational instability towards a BH. Complementary to this, we study the dynamics of the G-stars, adopting the more recent dataset for G2 by \citet{2021ApJ...923...69P}, rather than the previous one presented in \citet{2019ApJ...871..126G}. Since only the latter was used in \citet{2020A&A...641A..34B} to test the fermion-core BH alternative, it is of central interest here to reanalyze the new G2 data along with the S2 star. \\
\indent
The outline consists of a brief overview of the fermionic RAR-DM halo model in section \ref{sec:rar}; the study of geodesics in these geometries in section \ref{sec:orbits}; in section \ref{sec:mcmc} we present the methodology regarding the statistical comparison between the models; while in section \ref{sec:results} we present the results corresponding to the S-2 star (subsection \ref{subsec:s2}) and G-sources (subsection \ref{subsec:gobjects}). In subsection \ref{subsec: halo} we present a robustness analysis of the fermionic model by studying how much the DM core predictions change when slightly changing the outer halo boundary conditions, with special attention to the GAIA DR3 rotation curve data. We outline the conclusions in section \ref{sec:conclusions}.

\section{Model overview} 

\subsection{Fermionic \textit{core-halo} profiles} \label{sec:rar}

Self-gravitating systems of neutral fermions, when analyzed through a kinetic theory coupled to gravity, have been shown to provide an alternative scenario to that of N-body simulations, for the formation, equilibrium, and stability of DM halos \citep{2015PhRvD..92l3527C,2021MNRAS.502.4227A,2025arXiv250310870K}. 
This includes a successful explanation of the overall rotation curve of the Galaxy in terms of fermionic core-halo profiles, when constrained by both: (i) recent GAIA DR3 rotation curve data at $\sim 10^1$ kpc scales, and (ii) mpc-scale observations of the best-resolved S-cluster stars \citep{2025arXiv250310870K}.
These approaches correspond to the attempt to explain the DM halos by first principles of statistical mechanics and thermodynamics, superseding the early works on violent relaxation by \citet{lynden1967statistical}. Indeed, it has been shown that such DM halo fermionic solutions have a finite mass, do not suffer from the gravothermal catastrophe, correspond to a true statistical equilibrium state (maximum entropy state), and agree with observations \citep{2021MNRAS.502.4227A,2023ApJ...945....1K,2025arXiv250310870K}. By applying a maximum entropy principle onto such self-gravitating systems, it was found that the most likely coarse-grained distribution function at relaxation is of Fermi-Dirac type, including particle evaporation effects \citep{1998MNRAS.300..981C,CHAVANIS200489,2021MNRAS.502.4227A,2025arXiv250310870K}:
\begin{equation}
    f(r,\epsilon \leq \epsilon_c)=\frac{1-e^{\left[\epsilon-\epsilon_c(r)\right]/kT(r)}}{e^{\left[\epsilon-\mu(r)\right]/kT(r)}+1} \qquad f(r,\epsilon > \epsilon_c)=0
    \label{eqn:DF}
\end{equation}
where $\epsilon=\sqrt{c^2p^2+m^2c^4}-mc^2$ is the particle
kinetic energy, $\mu(r)$ and $\epsilon_c(r)$ are the chemical potential and the escape energy (with the particle rest-energy subtracted off), $k$ is the Boltzmann constant, and $T(r)$ is the effective temperature.
For a given fermion mass $m$, the three free model parameters are the central temperature $\beta_0=kT_0/mc^2$, central degeneracy $\theta_0=\mu_0/kT_0$, and central cut-off energy $W_0=\epsilon_{c,0}/kT_0$. Given that self-gravitating systems of fermions admit a perfect fluid approximation \citep{PhysRev.187.1767}, the equilibrium states result by solving the Tolman-Oppenheimer-Volkoff equations for hydrostatic equilibrium, together with the Tolman and Klein thermodynamic equilibrium conditions \citep{1930PhRv...35..904T,1949RvMP...21..531K} and energy conservation along a geodesic. For model equations, see \citet{2018PDU....21...82A,2019IJMPD..2843003A,2020A&A...641A..34B,2024A&A...689A.194M}. In \citet{2021MNRAS.502.4227A}, these core-halo DM profiles have been shown to form and remain stable on cosmological time scales, while in \cite{2025EPJC...85L..753} it was shown the role of such fermionic DM candidates on the formation of the first DM halos in the high redshift Universe. A possible new paradigm for supermassive BH formation from the gravitational collapse of the high density DM cores has been presented in \cite{2024ApJ...961L..10A} and references therein. 

\citet{2023ApJ...945....1K} presented a detailed study of the properties and allowed morphologies of these fermionic density profiles when contrasted with a large data set of rotation curves. In this paradigm, depending on the degenerate state of the particles at the centre, they can lead to the formation of a dense core in the inner region (i.e. $\theta_0 \gtrsim 10$, core-halo family), or not (i.e. $\theta_0 \ll -1$, halo-only family), where the particles behave in a Boltzmannian regime in the latter case.
Remarkably, for fermionic core-halo solutions with fermion masses of $mc^2 \sim \mathcal{O}(100\ \rm{keV})$, the degenerate and compact DM cores may work as an alternative to the BH paradigm at the centre of galaxies \citep{2018PDU....21...82A,2019IJMPD..2843003A,2024A&A...689A.194M}, or eventually collapse into one providing a novel BH formation mechanism from DM \citep{2021MNRAS.502.4227A,2023MNRAS.523.2209A,2024ApJ...961L..10A}. 

A solution of the RAR-DM model for the case of the Milky Way, with specific boundary conditions that agree with the overall rotation curve, the orbits of the 17 best resolved S-cluster stars and the G2 object, was presented in \citet{2020A&A...641A..34B, 2021MNRAS.505L..64B} for fermions of $56$~keV. 
More recently, fermionic solutions of this kind were applied to a different halo tracer, i.e., stellar streams \citep{2024A&A...689A.194M,Collazo2025arXiv250515550C}, extending to larger particle masses of about $300$ keV \citep{2024A&A...689A.194M}.

\begin{figure}
    \centering
    \includegraphics[width=\columnwidth]{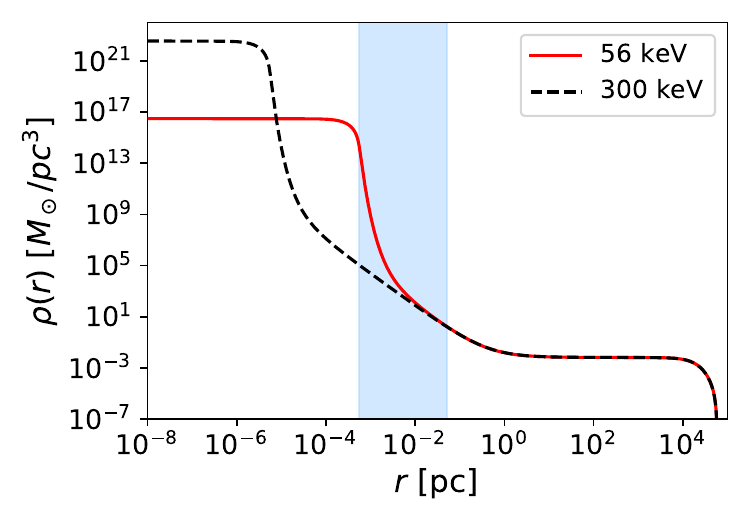}
    \caption{RAR-DM energy density profiles for fermions of $mc^2=56$ and $300$ keV for the same boundary conditions for the MW halo and core mass $M_{\rm c}$. The core compactness increases with the fermion mass. The blue band corresponds to the location of the most relevant S-stars and G-objects.}
    \label{fig:rhos}
\end{figure}

In figure \ref{fig:rhos}, we show two RAR density profiles of the core-halo family for fermions of $mc^2=56$ keV and $mc^2=300$ keV, satisfying the same halo mass constraints, (see Table \ref{tab:rar}) which agree with the Milky Way rotation curve as analysed in \citep{2018PDU....21...82A}.
The core mass $M_{\rm c}$ remains the same for both configurations, and it is defined at the maximum of the circular velocity curve (see figure \ref{fig:rot_curve}). Their corresponding locations are achieved approximately at a $12\%$ drop in central energy density $\rho_0$. As seen from figure \ref{fig:rhos}, 
the core compactness increases with the fermion mass, remaining completely inside the distance to the pericentres of S-cluster stars and G objects, represented by the blue band.

\subsection{Orbital dynamics} \label{sec:orbits}

The spherically symmetric spacetime metric can be written as
\begin{equation}
    \label{eqn:metric}
    ds^2 = A(r)c^2 dt^2 - B(r)dr^2 - r^2\left(d\vartheta^2 + \sin^2\vartheta d\phi^2\right)
\end{equation}
where $c$ is the speed of light, $A(r)$ and $B(r)$ are the metric functions to be found numerically by solving the Einstein field equations sourced by the fermionic perfect fluid system with the properties mentioned in the previous section \ref{sec:rar}. In the absence of a matter source, such equations can be solved analytically, leading to the Schwarzschild BH metric potentials
\begin{equation}
    A_{\rm BH}(r)=1-\frac{2G\MBH}{c^2r} \quad ; \quad B_{\rm BH}(r)=\frac{1}{A_{\rm BH}(r)}
\end{equation}
where $G$ is the gravitational constant and $\MBH$ the BH mass.

The equations of motion of a test particle in the space-time metric given by \cref{eqn:metric}, assuming without loss of generality the motion on the plane $\vartheta= \pi/2$, are given by 
\begin{subequations}\label{eqn:motion}
    \begin{eqnarray}
        \dot{t} &=& \dfrac{E}{c^2 A(r)},\label{eqn:motiont}\\
        \dot{\phi} &=&  \dfrac{L}{r^2},\label{eqn:motionphi}\\
        \ddot{r} &=& \dfrac{1}{2 B(r)}\left[- A'(r) c^2 \dot{t}^2 - B'(r) \dot{r}^2 + 2 r\dot{\phi}^2\right],\label{eqn:motionr}
    \end{eqnarray}
\end{subequations}
where $E$ and $L$ are the conserved energy and the angular momentum of the test particle per unit rest-mass, the over-dot stands for derivative relative to the proper time $\tau$, and the superscript comma ($'$) denotes the derivative relative to the radial coordinate $r$.
Specifying values for $E$ and $L$ --whose dependence is only on $A(r)$ at pericentre and apocentre-- we perform the numerical integration of  \cref{eqn:motion} with a Dormand-Prince algorithm  \citep{https://doi.org/10.1002/zamm.19880680638}, implemented as {\sc{DOP853}} by the Python library {\sc{SCIPY}}~\citep{2020SciPy-NMeth}. In addition, we choose appropriate initial conditions in such a way that the test particle motion starts at the apocentre, i.e., $t(\tau_{\rm 0}) = 0$, $\phi(\tau_{\rm 0})=\pi$, $r(\tau_{\rm 0})=r_{\rm a}$, and $\dot{r}(\tau_{\rm 0}) = 0$. 
Then, to compare the model with orbital data, we project the real orbit onto the plane of the sky. For this, we use the classic Thiele-Innes parameters \citep{2010arXiv1008.3416C} and account for the R{\o}mer time delay. Appendix C of \citet{2020A&A...641A..34B} shows a detailed description of how to implement this transformation.
In the BH scenario, once the central mass $\MBH$ has been specified, the metric potentials $A(r)$ and $B(r)$ are known. In the fermionic core-halo scenario, once the set of model parameters ($\beta_0,\theta_0,W_0$) is given for a fixed fermion mass $m$, the metric potentials $A(r)$ and $B(r)$ are determined.
In section \ref{sec:results}, we sample the parameter spaces via MCMC, maximizing a likelihood constructed by S2 astrometric data and incorporating the halo mass constraints (Table \ref{tab:rar}) for the RAR-DM model, to constrain the model parameters (see section \ref{sec:mcmc}). 

\begin{table}
    \centering
    \caption{Halo constraints for the Milky Way RAR-DM profiles taken from \protect \citet{2013PASJ...65..118S} and \protect \citet{2014MNRAS.445.3788G}.}
    \begin{tabular}{|c|c|} \hline
        Radius [kpc] & Cumulate mass [$M_\odot$] \\ \hline
        $12$ & $3.6\times 10^{10}$ \\
        $40$ & $2.3\times 10^{11}$ \\ \hline
    \end{tabular}
    \label{tab:rar}
\end{table}

\subsection{MCMC statistics and model comparison} \label{sec:mcmc}

The Bayesian framework of data analysis provides probabilistic ways to quantify the evidence that data provide in support of one scientific hypothesis over another. 
This can be implemented using Bayes factors, where models provide practical ways for these scientific hypotheses \citep{Jeffreys1939-JEFTOP-5}. 
Given a model $\mathcal{M}_{\rm i}$, a dataset $\mathcal{D}$ and a parameter vector $\mathbf{\Theta}$, the Bayes theorem states
\begin{equation}
    p(\mathbf{\Theta}|\mathcal{D},\mathcal{M}_{\rm i}) = \frac{p(\mathcal{D}|\mathbf{\Theta},\mathcal{M}_{\rm i}) p(\mathbf{\Theta}|\mathcal{M}_{\rm i})}{p(\mathcal{D}|\mathcal{M}_{\rm i})}
\end{equation}
where $p(\mathbf{\Theta}|\mathcal{D},\mathcal{M}_{\rm i})$ is the posterior probability of the model, $p(\mathcal{D}|\mathbf{\Theta},\mathcal{M}_{\rm i}) \equiv \mathscr{L}(\mathcal{D}|\mathbf{\Theta},\mathcal{M}_{\rm i})$ is the likelihood function,  $p(\mathbf{\Theta}|\mathcal{M}_{\rm i}) \equiv \pi(\mathbf{\Theta})$ is called the prior and expresses our knowledge about the model before acquiring the data, and $p(\mathcal{D}|\mathcal{M}_{\rm i}) \equiv \rm E_i$ is the marginal likelihood (also referred to as the Bayesian Evidence) and it acts as a normalizing factor, where the likelihood is evaluated for every possible parameter value weighted by the prior 
\begin{equation}
    {\rm E_i} \equiv p(\mathcal{D}|\mathcal{M}_{\rm i}) = \int {\bf d\Theta}\ \mathscr{L}(\mathcal{D}|\mathbf{\Theta},\mathcal{M}_{\rm i}) \pi(\mathbf{\Theta})
\label{eqn:evidence}
\end{equation}
To get a qualitative idea of how well a model $\mathcal{M}_{\rm i}$ may fit the data when compared to model $\mathcal{M}_{\rm j}$, Bayes factors are computed (or, equivalently, its logarithm), which are a measure of relative evidence  
\begin{equation}
    \mathcal{B}_{\rm i,j} = {\rm ln\left( \frac{E_i}{E_j} \right)}.
    \label{eqn:bayes factor}
\end{equation}
A scale shown in Table \ref{tab:evidence} has been proposed to interpret Bayes factors according to the strength of evidence in favour of one model over another \citep{Jeffreys1939-JEFTOP-5}.

The log-likelihood function is given by 
\begin{equation}
   \text{ln} (\mathscr{L}) = -\frac{1}{2} \sum_{\rm k = 1}^{\rm n} \frac{(\rm{X_k} - \rm{X_{M,k}})^2}{\sigma_{\rm X_k}^2},
\end{equation}
where $\rm {X_k}$ are the elements of a vector containing the observed data, $\rm {X_{M,k}}$ are the predicted values for a given model $\mathcal{M}$; and $\sigma_{\rm X_k}$ are the associated observational uncertainties.
To apply this formalism, we selected the following models to test the S2 star dynamics:
\begin{enumerate}
    \item The \textit{null} model $\mathcal{M}_0$ corresponds to the traditional approach of a BH at distance $R_\odot$ and mass $\MBH$. The 10-dimensional parameter space vector $\mathbf{\Theta}$ is given by the 6 orbital parameters, plus 4 determining the gravitational potential: $\MBH$, $R_\odot$ and $X_{\rm off}$, $Y_{\rm off}$ that correspond to offsets in right ascension (RA) and declination (DEC), respect to the BH sky position.

    \item Model $\mathcal{M}_1$ is given by RAR-DM fermions of $mc^2 = 56$~keV, which result in a relatively low compactness of the central core. The 12-D parameter space vector $\mathbf{\Theta}$, is constituted by the 6 orbital parameters, $X_{\rm off}$ and $Y_{\rm off}$ positional offsets, the distance to the central core $R_\odot$ and the thermodynamic parameters $\beta_0$, $\theta_0$ and $W_0$. 

    \item Model $\mathcal{M}_2$ is given by RAR-DM fermions of $mc^2 = 300$~keV, resulting in a high compactness of the central core (see figure \ref{fig:rhos}). The parameter space $\mathbf{\Theta}$ is the same 12-dimensional vector as model $\mathcal{M}_1$.
\end{enumerate}

For the case of the G-objects, we fix the gravitational potentials produced by $\mathcal{M}_0$, $\mathcal{M}_1$, and $\mathcal{M}_2$, due to small orbital coverage and data quality. 
We name these simplified models as:

\begin{enumerate}
    \item $\mathcal{M}_{0a}$ corresponds to a BH with fixed potential parameters ($X_{\rm off}$, $Y_{\rm off}$, $R_\odot$, $M_\bullet$), given in Table \ref{tab:S2} for $\mathcal{M}_0$.
    The parameter space vector $\mathbf{\Theta}$ is 6-dimensional, given only by the orbital parameters.
    
    \item $\mathcal{M}_{1a}$ corresponds to fermions of $mc^2=56$ keV with fixed potential parameters ($X_{\rm off}$, $Y_{\rm off}$, $R_\odot$, $\beta_0$, $\theta_0$, $W_0$), given in Table \ref{tab:S2} for $\mathcal{M}_1$. 
    Likewise, the parameter space vector $\mathbf{\Theta}$ is 6-dimensional.
    
    \item $\mathcal{M}_{2a}$ corresponds to fermions of $mc^2=300$ keV with fixed potential parameters ($X_{\rm off}$, $Y_{\rm off}$, $R_\odot$, $\beta_0$, $\theta_0$, $W_0$), given in Table \ref{tab:S2} for $\mathcal{M}_2$. 
    Likewise, the parameter space vector $\mathbf{\Theta}$ is 6-dimensional.
\end{enumerate}

The datasets $\mathcal{D}$ for these stars are constituted by RA and DEC positions in the plane of the sky plus radial velocity, taken from \citet{2019Sci...365..664D,2020A&A...634A..35P,2020Natur.577..337C}.
We assumed uniform priors in every case. 
For the parameter space exploration, we use a statistical sampler with an affine-invariant Markov Chain Monte Carlo (MCMC) method implemented in Python3 (\texttt{emcee} package \citealp{2013PASP..125..306F}). 
The convergence of the MCMC analysis is guaranteed by the \textit{integrated auto-correlation time} $\tau_{\rm f}$, which directly quantifies the Monte Carlo error. We seek chains long enough that for N numbers of samples, it satisfies $ \rm N \gg 50\ \tau_{\rm f}$. For the estimation of the Bayes factors \cref{eqn:bayes factor} for every model, we use \texttt{MCEvidence}, a Python package that implements marginal likelihoods from the MC chains \citep{heavens2017marginallikelihoodsmontecarlo}.

\begin{table}
    \caption{Bayes factor scale for model comparison as proposed by \protect \citet{Jeffreys1939-JEFTOP-5}}
    \centering
    \normalsize{
    \begin{tabular}{lccr}
    \hline
    $|\mathcal{B}_{\rm i,j}|$ & odds & probability & strength \\ \hline
    $<1.0$ & $<3 : 1$ & $<0.750$ & inconclusive \\
    $1.0-2.5$ & $\sim 12 : 1$ & $<0.923$ & significant \\
    $2.5-5.0$ & $\sim 150 : 1$ & $<0.993$ & strong \\
    $>5.0$ & $>150 : 1$ & $>0.993$ & decisive \\ \hline
    \end{tabular}
    }
    \label{tab:evidence}
\end{table}

\section{Results} \label{sec:results}

In this section we apply the formalisms described above for the S2 star data \citep{2019Sci...365..664D}, which most strongly constrains the gravitational potential of SgrA* due to the quantity and quality of data, and 5 stars of the G-cluster \citep{2020A&A...634A..35P,2020Natur.577..337C}. 

\subsection{Results on S2 star} \label{subsec:s2}

The most relevant of the S-cluster stars is the S2 star, completing an orbit in $\sim 16$ years around the Galactic Centre in a highly eccentric orbit $e \sim 0.88$ and is the brightest source of light from the whole S-star cluster $m_{\rm K} \sim 14$ \citep{2018A&A...615L..15G,2019Sci...365..664D}. 
To corroborate and extend the results obtained in \citet{2020A&A...641A..34B} for the specific case of $mc^2=56$ keV fermions, the orbital analysis as presented in section \ref{sec:orbits} was done for diverse gravitational potentials represented by a BH model ($\mathcal{M}_0$) and RAR-DM fermionic models ($\mathcal{M}_1$ and $\mathcal{M}_2$). The sampling of the space parameters was performed with MCMC technique for the S2 star data taken from \citet{2019Sci...365..664D}. 

\begin{figure}
    \centering
    \includegraphics[width=\columnwidth]{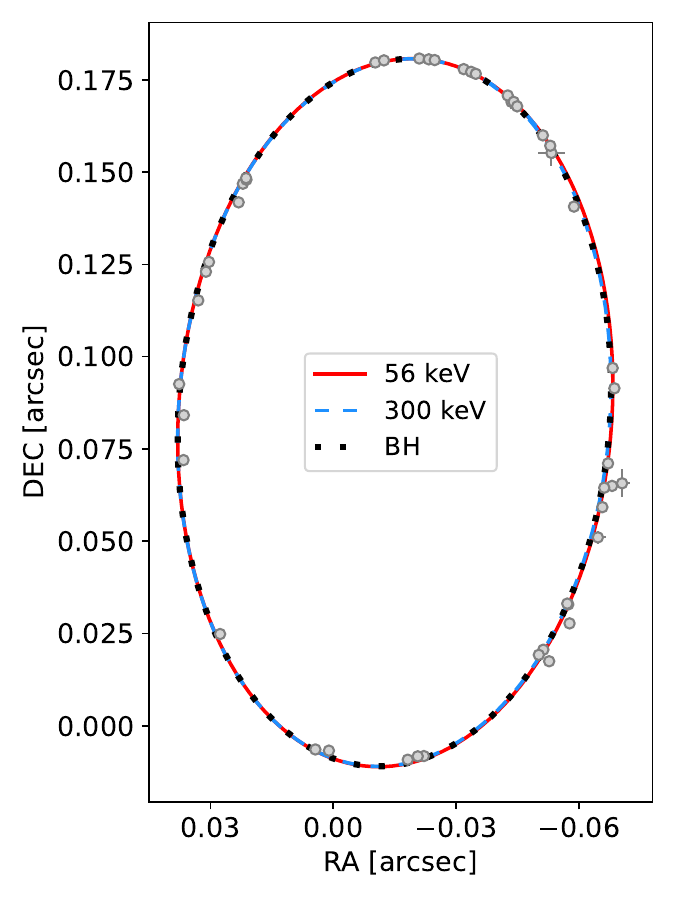}
    \includegraphics[width=\columnwidth]{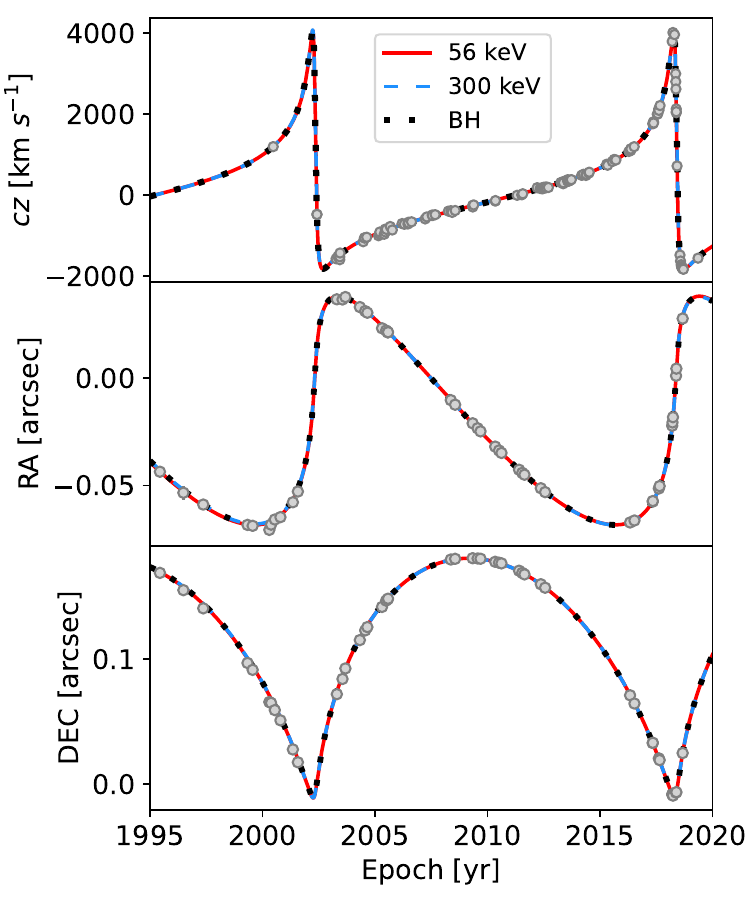}
  \caption{\textit{Top:} Observed and theoretical orbits of S2 star. \textit{Bottom:} radial velocity, Right Ascension, and Declination as a function of time. The astrometric measurements were taken from \citet{2019Sci...365..664D}. For a better visualization of the orbit at the apocentre and pericentre, see figure 2 in \citet{2022MNRAS.511L..35A}.}
  \label{fig: S2_orbit}
\end{figure}

In contrast to the $\mathcal{M}_0$ model (i.e., the BH) that only fits the gravitational potential below $\sim$ pc scale, the fermionic halos aim to model not only the potential on this scale, but also the DM halo component of the Galaxy. Hence, the dataset used for models $\mathcal{M}_1$ and $\mathcal{M}_2$ need to add (at least) two observationally inferred enclosed mass values on halo scales (given in Table \ref{tab:rar}) in addition to the photometric and spectroscopic orbital data of the S2 star. Strictly speaking, we do not have the same likelihood (due to different datasets), so a comparison with the \textit{null} model via Bayes factors would not be properly applied. 
Table \ref{tab:S2} compares the best-fit values for the parameter spaces along with model-predicted quantities. The table shows that the accuracy of the data implies inferred parameters with relative errors lower than $1\%$ among the three models, resulting in nearly identical trajectories for the reconstructed orbits (see figure \ref{fig: S2_orbit}). These parameters also agree with the values derived by other works \citep{2017ApJ...837...30G,2017ApJ...845...22P,2019Sci...365..664D,2020A&A...641A..34B,2020A&A...636L...5G}. 

Among the model predicted quantities, we derive central masses of $M_{\rm BH}=4.16\times10^6 M_\odot$ for a central BH ($\mathcal{M}_0$), $M_{\rm c}=3.52\times10^6 M_\odot$ for a $56$ keV fermionic core ($\mathcal{M}_1$) and $M_{\rm c}=3.54\times10^6 M_\odot$ for a 300 keV fermionic core ($\mathcal{M}_2$), denoted with $M_\bullet$ row in Table \ref{tab:S2}. However, in the fermionic models the total mass enclosed by the S2 pericentre ($M(r_{\rm p})$ row in Table \ref{tab:S2}) corresponds to $M(r_{\rm p})=4.05\times10^6 M_\odot$ for $\mathcal{M}_1$ and $M(r_{\rm p})=4.16\times10^6 M_\odot$ for $\mathcal{M}_2$, in agreement with the central mass required to explain the S2 dynamics. 
The predicted positions in the plane of the sky, the radial velocity, and data are shown in figure \ref{fig: S2_orbit}. 
We perform, however, a statistical comparison of the fermionic models $\mathcal{M}_1$ and $\mathcal{M}_2$, which do share the same likelihood function. The Bayes factor shown in Table \ref{tab:Bayes-factors} gives a value of $\mathcal{B}_{1,2}=10.1$, implying evidence (when using this dataset) in favour of the potential generated by a $56$ keV fermionic DM model against a $300$ keV DM model. However we emphasize that error bars in the S2 astrometric data from \cite{2019Sci...365..664D} at about pericentre passage, has been considerably reduced after the GRAVITY instrument \cite{2022A&A...657L..12G}, and thus our conclusions still need to be verified in light of such new data. 
In figures \ref{fig:56keV cornerplot} and \ref{fig:300keV cornerplot}, we show the marginalized distributions for models $\mathcal{M}_1$ and $\mathcal{M}_2$. 

\subsection{Results on G objects} \label{subsec:gobjects}

Motivated by the new orbital data on the G2 object reported by \citet{2020A&A...634A..35P,2023ApJ...943..183P}, along with other 4 stars reported by \citet{2020Natur.577..337C}, we aim to study the dynamics of these stars under the gravitational potentials described in section \ref{sec:mcmc}: $\mathcal{M}_{0a}$, $\mathcal{M}_{1a}$ and $\mathcal{M}_{2a}$. Given the low orbital coverage collected so far of G-objects, we adopt fixed central gravitational potentials inferred from the S2 orbital data.

\begin{figure*}
  \centering
  \includegraphics[scale=0.645]{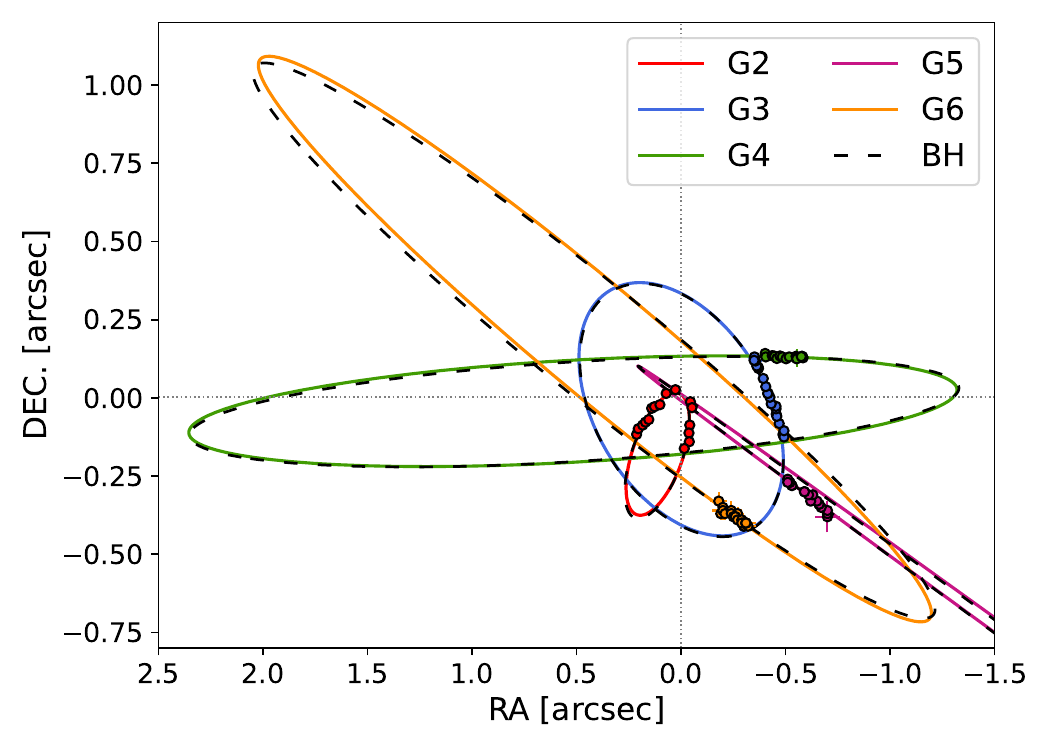}
  \caption{Observed and theoretical orbits for the G-cluster stars. coloured lines correspond to the best-fit orbits modelled with $\mathcal{M}_{1a}$ of 56 keV fermions. Dashed black lines correspond to the best-fit orbits modelled with $\mathcal{M}_{0a}$ a BH. The astrometric measurements are taken from \citet{2020Natur.577..337C,2023ApJ...943..183P}.}
  \label{fig:G-orbits}
\end{figure*}

\begin{figure*}
  \centering
   \includegraphics[width=\textwidth]{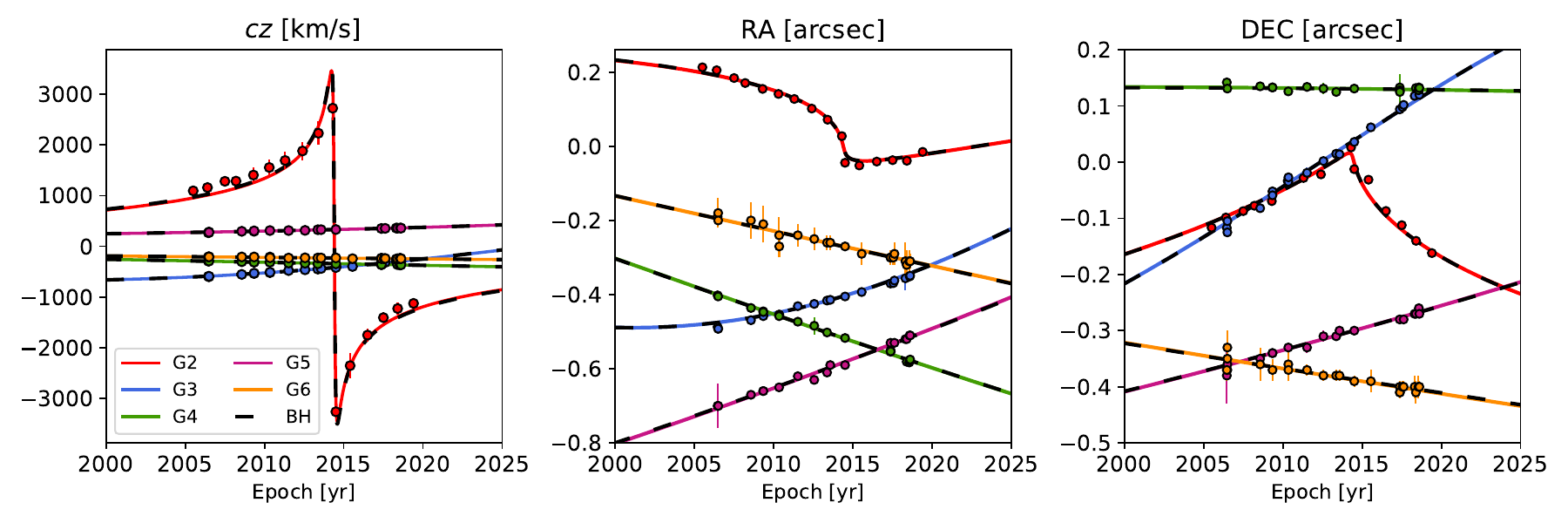}
  \caption{Observed and theoretical right ascension, declination, and radial velocity for the G-cluster stars. The astrometric measurements are taken from \citet{2020Natur.577..337C,2023ApJ...943..183P}. As in figure \ref{fig:G-orbits}, coloured lines correspond to model $\mathcal{M}_{1a}$, and black dashed lines to model $\mathcal{M}_{0a}$.}
  \label{fig:G-data}
\end{figure*}

In figure \ref{fig:G-orbits}, we show the projected orbits in the plane of the sky for the 5 G-objects, modelled with $\mathcal{M}_{0a}$ in dashed black lines and select only $\mathcal{M}_{1a}$ in solid coloured lines, for better clarity in the graphs. The same for figure \ref{fig:G-data} where we show the RA., DEC., and line of sight velocity $cz$ as a function of time. The inferred orbital parameters are shown in Table \ref{tab:G-parameters} for all models.
For these \textit{fixed} models, we perform a qualitative comparison using data from the G-objects, by computing Bayes factors \cref{eqn:bayes factor} and presenting them in Table \ref{tab:Bayes-factors}. When comparing models $\mathcal{M}_{0a}$ and $\mathcal{M}_{2a}$, we found Bayes factors of $\mathcal{B}< 1$ for all G-stars, indicating that the data does not favour the \textit{null} model over the potential of a $300$ keV fermionic core. The same conclusion can be drawn for a fermionic core of 56 keV, when contrasted with G4, G5, and G6 data. Instead, when comparing models $\mathcal{M}_{0a}$ and $\mathcal{M}_{1a}$ with the G2 orbital data we found significant evidence in favour of the \textit{null} model ($\mathcal{B}_{0a,1a}=2.18$); and strong evidence in favour of the \textit{null} model in the case of the G3 orbital data ($\mathcal{B}_{0a,1a}=3.93$).

\begin{table}
\caption{Bayes factors of fermionic DM models compared to the null model ($0a$) of a BH. Subscript $j$ corresponds to $1a$ or $2a$. In order to statistically compare we fixed the gravitational potentials of the best fit of models with S2 constraints (maximum likelihoods, Table \ref{tab:S2}), and re-sampled over the orbital parameters for G objects. The last row corresponds to the Bayes factor comparing models $\mathcal{M}_1$ and $\mathcal{M}_2$ for the S2 astrometric dataset plus MW halo.}
\centering
\normalsize{
\begin{tabular}{|l|cc|r|}
\hline
 & $\mathcal{M}_{1a}$ & $\mathcal{M}_{2a}$ & \\ \hline 
\multicolumn{1}{|l|}{} & $2.18$ & $0.29$ & \multicolumn{1}{r|}{G2} \\ \cline{2-4} 
\multicolumn{1}{|l|}{} & $3.93$ & $0.15$ & \multicolumn{1}{r|}{G3} \\ \cline{2-4} 
\multicolumn{1}{|l|}{$\mathcal{B}_{\rm 0a,j}$} & $-0.36$ & $-0.03$ & \multicolumn{1}{r|}{G4} \\ \cline{2-4} 
\multicolumn{1}{|l|}{}  &$0.59$ & $0.86$ & \multicolumn{1}{r|}{G5} \\ \cline{2-4} 
\multicolumn{1}{|l|}{}  & $0.20$ & $0.06$ & \multicolumn{1}{r|}{G6} \\ \hline \hline
 & $\mathcal{M}_{1}$ & $\mathcal{M}_{2}$ & \\ \hline 
$\mathcal{B}_{\rm 1,2}$ & \multicolumn{2}{c|}{$10.1$} & S2\\ \hline 
\end{tabular}
}
\label{tab:Bayes-factors}
\end{table}
%

\subsection{Robustness of the results} \label{subsec: halo}

To assess the robustness of the DM core predictions in the fermionic models (i.e. as derived for the S2 star in section \ref{subsec:s2} with models $\mathcal{M}_1$ and $\mathcal{M}_2$), we performed an equivalent statistical analysis for the simultaneous fit of the S2 star orbital data along with slightly different DM halo boundary conditions. That is, we now consider new DM halo mass and radii of $M(10\ {\rm kpc})= 4.8\times 10^{10} M_\odot$ and $M(30\ {\rm kpc})= 1.4\times 10^{11} M_\odot$ as analysed in the context of the fermionic model in \citet{2025arXiv250310870K}, instead of the ones given in Table \ref{tab:rar}. 
This election corresponds to a mass profile consistent with the most recent GAIA DR3 \citep{2023A&A...674A...1G} rotation curve data. Our aim is to test how sensitive are the fermionic core potentials, as e.g. for the particle mass $mc^2=300$ keV, to different DM halo constraints. 

Analogously to what was done in section \ref{sec:mcmc}, we performed a parameter space exploration with the MCMC technique for the 12-parameter spaces, evaluating the new likelihood function. The sampling projected distributions are plotted in figure \ref{fig:300keV cornerplot GAIA} for the case of $mc^2=300$ keV fermions. 
We thus confirm that the new DM halo boundary conditions, taken from GAIA-DR3 data, which introduce modifications in the halo constraint amounting to $30\%$ in the inferred outer mass profile $M(r)$, have a remarkably unchanged impact on the orbital parameters of S2. The relative variation in all fitted orbital parameters does not exceed $0.1\%$, indicating the robustness of the fermionic models in predicting stellar orbits. 

\begin{figure}
    \centering
    \includegraphics[width=\columnwidth]{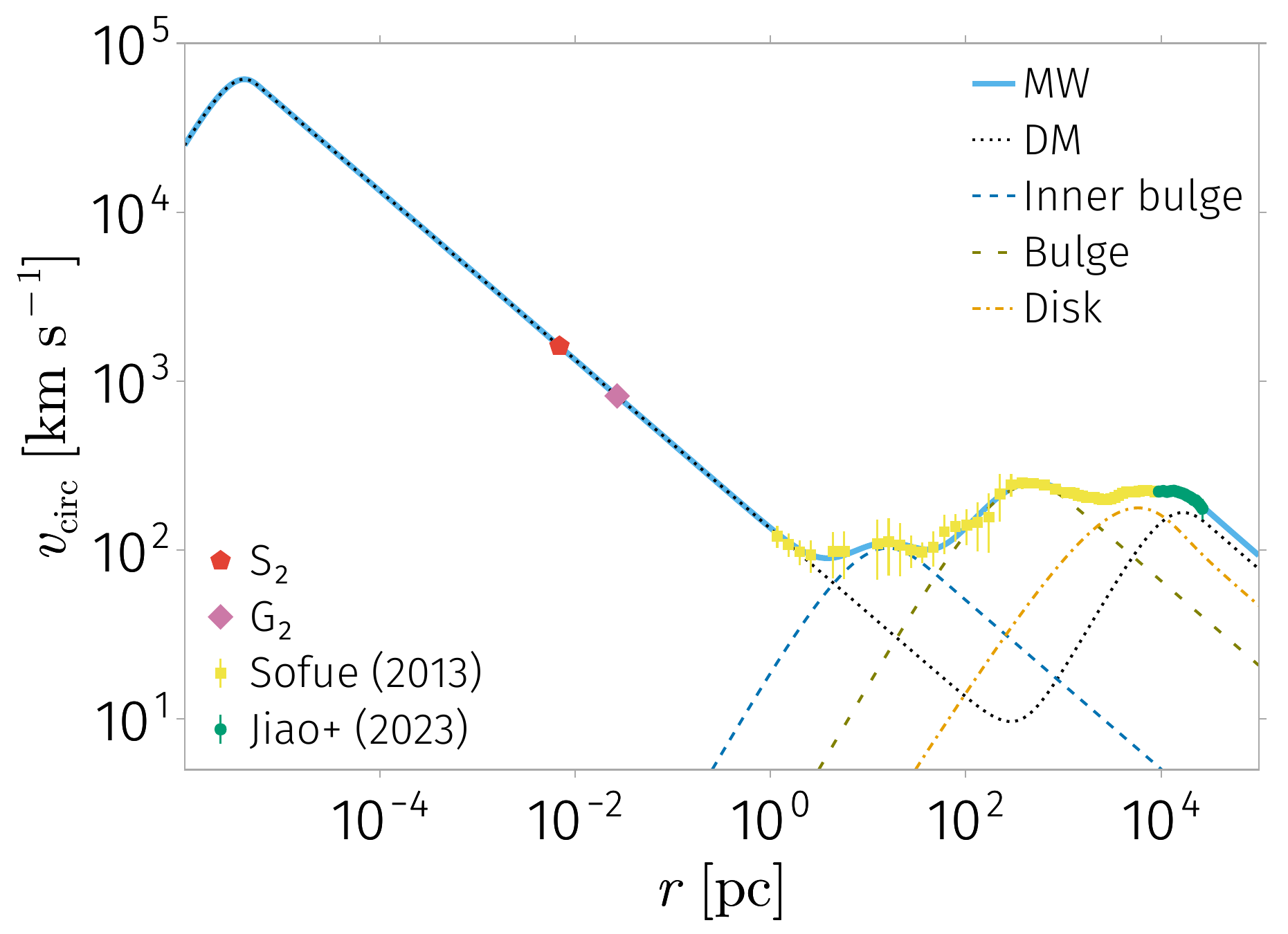}
    \caption{MW Rotation curve. Data and models reads: S2 (red pentagon), G2 (violet diamond), Inner Bulge (dashed blue line), Bulge (dashed olive line), Disk (dot-dashed orange line), fermionic DM with $mc^2=300$ keV (dotted black line), full MW (solid light-blue line) and halo data are taken from \citet{sofue_rotation_2013} (yellow squares) and \citet{Jiao2023} (green circles).}
    \label{fig:rot_curve}
\end{figure}

\begin{figure}
    \centering
    \includegraphics[width=\columnwidth]{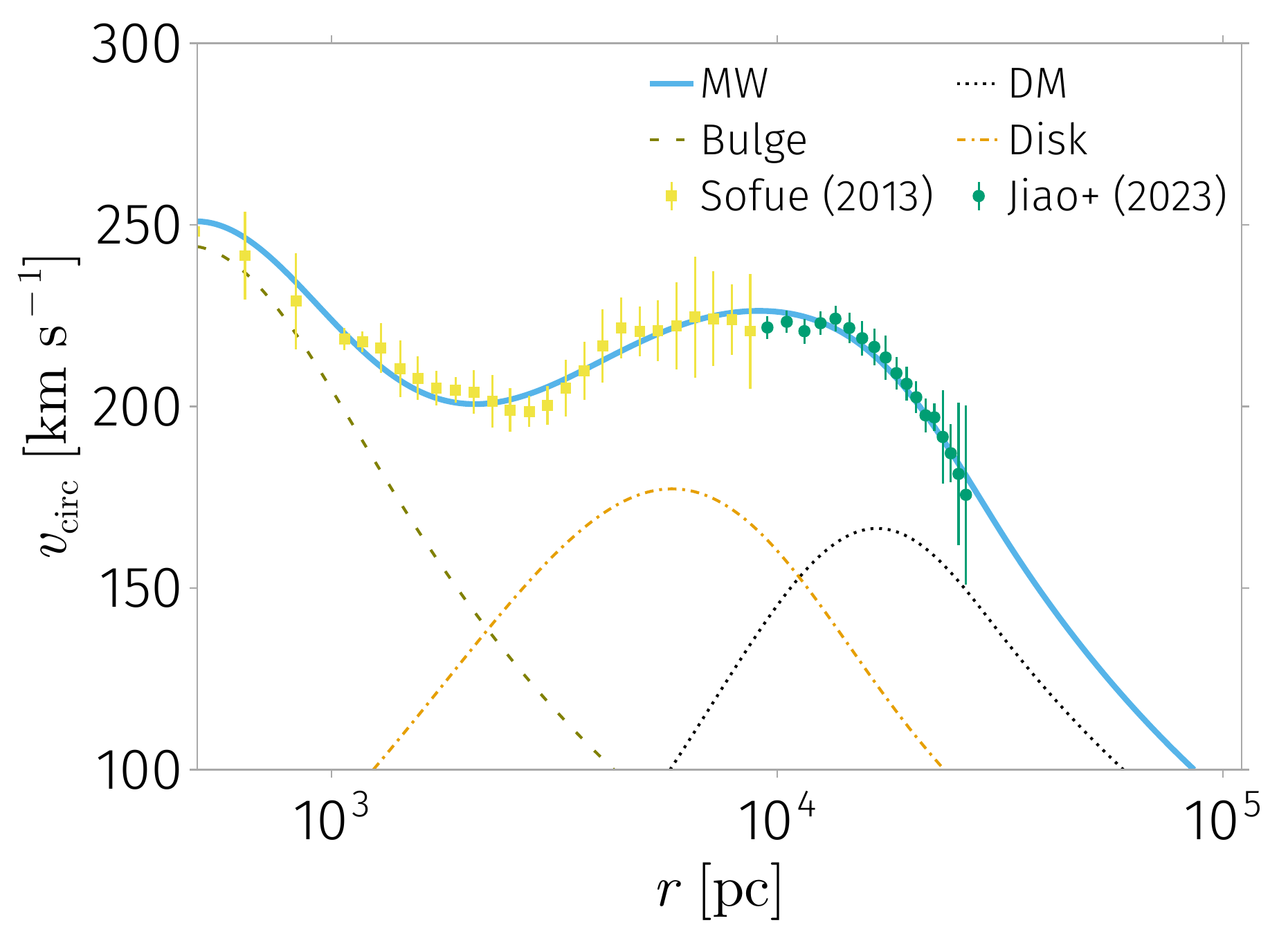}
    \caption{MW rotation curve on intermediate-to-outer halo scales. Data and models are displayed as in figure \ref{fig:rot_curve}.}
    \label{fig:rot_curve_halo}
\end{figure}

Finally, we show in figures \ref{fig:rot_curve} and \ref{fig:rot_curve_halo} the total and halo-scale MW rotation curve, with data adopted from \citet{sofue_rotation_2013,Jiao2023} and a mean circular velocity value calculated for S2 and G2 stars at the mpc scale. The DM model considered corresponds to the best-fit for the S2 star together with GAIA-DR3 cumulated mass constraints given above for a $mc^2=300$ keV fermionic potential, while the baryonic components (i.e., disc and bulge) are computed following the mass models given in \citep{2025arXiv250310870K}.

\begin{table}
\caption{Summary of the inferred best-fit orbital parameters of models $\mathcal{M}_0$ (BH), $\mathcal{M}_1$ ($56$ keV fermions), $\mathcal{M}_2$ ($300$ keV fermions) for the S2 star. 
Model-predicted quantities are shown at the bottom.
$1\sigma$ uncertainties can be seen in figures \ref{fig:56keV cornerplot},\ref{fig:300keV cornerplot}.}
\centering
\normalsize{
\begin{tabular}{llccc}
\hline
\multicolumn{2}{l}{Parameter} & $\mathcal{M}_1$ & $\mathcal{M}_2$ & $\mathcal{M}_0$\\
\hline 
$a$ & ({as})  & $0.1252$ & $0.1246$ & $0.1247$ \\
$e$ & & $0.8864$ & $0.8854$ & $0.8854$\\
$t_{\rm p}$ & (yr)  & $2018.38$ & $2018.38$ & $2018.38$\\
$\omega$ & ($^{\circ}$)  & $66.783$ & $66.394$ & $66.403$\\
$i$ & ($^{\circ}$)  & $134.44$ & $134.70$ & $134.67$\\
$\Omega$ & ($^{\circ}$)  & $228.11$ & $228.02$ & $228.03$\\
$X_{\rm off}$ & (mas)  & $-0.061$ & $-0.072$ & $-0.094$\\
$Y_{\rm off}$ & (mas)  & $2.4337$ & $2.5739$ & $2.5918$\\ 
$M_{\bullet}$ & $(\times 10^6 M_\odot)$  & $3.5156$ & $3.5388$ & $4.1563$\\ 
$R_{\odot}$ & (kpc) & $8.1124$ & $8.2126$ & $8.2064$ \\ 
$\beta_{0}$ & & $1.21\times 10^{-5}$ & $1.31\times 10^{-3}$ & -- \\ 
$\theta_0$ & & $37.571$ & $42.280$ & -- \\ 
$W_0$ & & $66.005$ & $72.896$ & -- \\ \hline
$P$ & (yr)  & $16.048$ & $16.052$ & $16.051$\\
$r_{\rm p}$ & ({as})  & $0.0142$ & $0.0143$ & $0.0143$\\
$r_{\rm a}$ & ({as})  & $0.2361$ & $0.2350$ & $0.2350$\\
$\Delta \phi$ & (min rev$^{-1}$)  & $-2.053$ & $12.048$& $12.035$\\ 
$M(r_{\rm p})$ & ($\times 10^6 M_\odot$)  & $4.0464$ & $4.1619$ & $4.1563$\\
$\Delta M$ & ($M_\odot$)  & $1.95\times 10^{4}$ & $1.52\times 10^{-3}$ & -- \\ \hline
\end{tabular}
}
\begin{tablenotes}%
\item corresponding to $a$: semi-major axis; $e$: eccentricity; $t_{\rm p}$: epoch of pericentre passage; $\omega$: argument of pericentre; $i$: inclination; $\Omega$: position angle of the ascending node; $X_{\rm off}$, $Y_{\rm off}$: offsets of the central object in RA. and DEC. respectively; $M_\bullet$: $\MBH$ in $\mathcal{M}_0$, core mass (model predicted) in $\mathcal{M}_{1,2}$; $R_\odot$: Sun's distance to Sgr A*; $\beta_0$: DM central temperature ; $\theta_0$: DM central degeneracy; $W_0$: DM central cut-off energy; $P$: orbital period; $r_{\rm p}$: distance to pericentre; $r_{\rm a}$: distance to apocentre; $\Delta \phi$: precession angle per orbital period; $M(r_{\rm p})$: enclosed mass in the pericentre; $\Delta M$: extended DM mass between pericentre and apocentre.
\end{tablenotes}
\label{tab:S2}
\end{table}

\newpage

\begin{table*}
\caption{Summary of the inferred best-fit orbital parameters for $\mathcal{M}_{0a}$ and $\mathcal{M}_{1a}$ corresponding to figures \ref{fig:G-orbits},\ref{fig:G-data}. \\
Model-predicted quantities are separated at the bottom. $1\sigma$ uncertainties can be seen in figures \ref{fig:G2_MCMC},\ref{fig:G3_MCMC}.} 
\label{tab:G-parameters}
\centering
\resizebox{\textwidth}{!}{%
\begin{tabular}{clccccccccccccccc} 
\hline
\multicolumn{2}{c}{\multirow{2}{*}{Parameter}}&  & \multicolumn{2}{c}{G2}&  & \multicolumn{2}{c}{G3}&  & \multicolumn{2}{c}{G4}&  & \multicolumn{2}{c}{G5}&  & \multicolumn{2}{c}{G6} \\ 
\cline{4-5} \cline{7-8} \cline{10-11} \cline{13-14} \cline{16-17} 
\multicolumn{2}{c}{}&  & $\mathcal{M}_{0a}$ & $\mathcal{M}_{1a}$&  & $\mathcal{M}_{0a}$ & $\mathcal{M}_{1a}$ &  & $\mathcal{M}_{0a}$ & $\mathcal{M}_{1a}$ &  & $\mathcal{M}_{0a}$ & $\mathcal{M}_{1a}$ &  & $\mathcal{M}_{0a}$ & $\mathcal{M}_{1a}$ \\ 
\cline{1-2} \cline{4-4} \cline{5-5} \cline{7-7} \cline{8-8} \cline{10-10} \cline{11-11} \cline{13-13} \cline{14-14} \cline{16-16} \cline{17-17}   
$a$ &({as}) & & {$0.464$} & {$0.450$} & & {$0.539$} & {$0.535$} & & {$1.869$}& {$1.865$}& & {$2.270$}& {$2.366$}& & {$1.881$}& {$1.872$}\\
$e$ & & &{$0.970$} & {$0.969$} & & {$0.110$}& {$0.102$}& & {$0.321$}& {$0.324$}& & {$0.900$} & {$0.904$}& & {$0.306$}& {$0.309$}\\
$t_{\rm p}$ & (yr)& & {$2014.41$} & {$2014.41$}& & {$2035.44$}& {$2035.80$}& & {$2060.63$}& {$2063.17$}& & {$2062.01$}& {$2061.57$} & &{$2226.45$} &{$2224.69$}\\
$\omega$ & ($^{\circ}$) & & {$91.55$} & {$91.71$}& & {$293.81$} & {$294.78$}& & {$148.70$} & {$149.34$}& & {$9.82$}& {$9.18$} & &{$218.14$} & {$218.79$}\\
$i$ & ($^{\circ}$)& & {$119.31$}& {$119.31$}& & {$51.15$}& {$51.41$}& & {$95.10$} & {$95.23$} && {$91.12$} & {$91.23$}& & {$83.61$} & {$83.74$}\\
$\Omega$ & ($^{\circ}$) & & {$61.73$}& {$61.81$} & & {$57.80$}& {$57.95$} && {$92.13$}& {$91.90$}& & {$64.15$} & {$64.29$}& & {$60.70$} &{$60.79$}\\ \hline
$P$ & (yr) && {$115.27$} & {$109.78$}& & {$144.18$}& {$142.36$}& & {$931.61$}& {$925.39$}& & {$1246.99$} & {$1322.51$}& & {$940.74$} & {$930.32$}\\
$r_{\rm p}$ & ({as}) & & {$0.014$}  & {$0.014$}& & {$0.479$}& {$0.481$}& & {$1.268$}& {$1.262$}& & {$0.225$} & {$0.227$} & & {$1.306$} &{$1.293$}\\
$r_{\rm a}$ & ({as}) & & {$0.877$}& {$0.886$} & & {$0.598$}& {$0.590$} & &{$2.469$} &{$2.468$} & & {$4.314$} & {$4.505$} & & {$2.456$} &{$2.450$}\\
$\Delta \phi$ & (min rev$^{-1}$) && {$11.72$}& {$-2.88$}& & {$0.61$} &{$0.60$} & &{$0.19$} & {$0.19$} & &{$0.76$}& {$0.74$} & &{$0.19$}& {$0.19$}\\
\hline
\end{tabular}}
\begin{tablenotes}%
\item where $a$: semi-major axis of the orbit; $e$: eccentricity; $t_{\rm p}$: epoch of pericentre passage; $\omega$: argument of pericentre; $i$: inclination; $\Omega$: position angle of the ascending node; $P$: orbital period; $r_{\rm p}$: distance to pericentre; $r_{\rm a}$: distance to apocentre; $\Delta \phi$: precession angle. 
\end{tablenotes}
\label{tab:G-cluster}
\end{table*}

\label{sec:rotation_curve}


\section{Discussion and conclusions} \label{sec:conclusions}

In the first part of this work, we studied the dynamics of the S2 star using the reduced orbital data from $2000$ to $2019$ in \citet{2019Sci...365..664D}. We tested the gravitational potential of Sgr A* when modelled with fermionic DM compact cores of different degrees of compactness, as proposed by the extended RAR model, against the traditionally assumed BH paradigm. 
We resort to Bayesian statistics to robustly constrain the parameter space exploration via MCMC, and to compare the different models assumed using Bayes factors. 
In the latter, a fair comparison between the potentials of the BH and fermionic models is not possible. The DM models have three additional parameters that not only adjust the potential in the centre but also in the outer halo of the Galaxy to explain the rotation curve. On the latter aspect, we gave special emphasis to the latest GAIA DR3 rotation curve data (see figure \ref{fig:rot_curve}).
Therefore, it is necessary to add (at least) two data constraints in the outer halo. Since we changed the dataset, we changed the likelihood function and did not perform a Bayes factor comparison. This, in turn, can be applied between DM potentials of different fermion masses.

We selected $mc^2=56$ keV illustrating a case of low compactness of the central core as previously studied in \citet{2020A&A...641A..34B,2021MNRAS.505L..64B,2022MNRAS.511L..35A} for the S2 star, and $mc^2=300$ keV illustrating a case of high core-compactness, which implies an enclosed DM mass in the S2 orbit of $\sim 10^{-8}$ times the one in the $56$ keV case.

Due to the relatively lower accuracy in astrometric data used (as, e.g., compared by GRAVITY data \cite{2022A&A...657L..12G}, though not publicly available in reduced form), the conclusive Bayesian evidence in favour of the $56$ keV fermionic model should be taken with caution.
Indeed, the use of a more accurate dataset could change this conclusion.
However, given the current lack of direct imaging data within the GRAVITY Collaboration (for example from NIRCAM/Keck or ERIS/VLT), we therefore adopted in this work an orbital solution that has been cross-validated by independent analyses and instruments, as done in \cite{2019Sci...365..664D} after 16 years of coverage using an imager with the highest spatial pixel scale available to date. \\
A difference between the models here considered is in the predicted amount of relativistic periapsis precession in two consecutive orbits. As shown in \citet{2022MNRAS.511L..35A}, already 
for fermions of $mc^2 \gtrsim 60$ keV, the precession angle begins to resemble the one predicted by a Schwarzschild BH (see also Table \ref{tab:S2}). 
Furthermore, in \citet{2020A&A...636L...5G}, a detection of $\sim 12$ arcmin/revolution was reported with the new increased accuracy of photometric data from the GRAVITY instrument. Once tested with a better resolution and larger dataset -- especially containing data from a future passage through the S2 apocentre in 2026 -- the new results are expected to disfavour the $56$ keV fermionic DM model, which predicts retrograde precession of $\sim -2$ arcmin/revolution.

Another argument disfavouring low particle masses (e.g., $56$ keV) over more compact fermion-core configurations is the stability analysis of fermionic profiles dedicated to the Milky Way in \citet{2025arXiv250310870K}, hence suggesting the fermion mass between $\sim$ ($200,378$) keV. Though even if low-compactness cores seem to be disfavoured by the above reasoning, there is yet a different way to interpret this result. Namely, the less compact-core solution could instead represent an effective mass distribution around a more compact core, analogously to the case of a BH surrounded by an extended mass, which could (potentially) explain the improvement in the orbital fitting.   

Another observable that favours fermionic DM compact-core solutions over ones with lower compactness is the image of Sgr A* presented in 2022 by the EHT collaboration \citep{EHTCollaboration_2022}. There, it was reported that the image of Sgr A* shows a central brightness depression with a diameter of $\sim 52\ \mu$as, consistent with a BH of mass $\sim 4\times 10^6 M_\odot$ and distance $R \approx 8$ kpc. Recently, in \citet{2024MNRAS.534.1217P}, it was shown that sufficiently high DM core compactness can reproduce similar images to those of BHs when surrounded by an accretion disk. For Milky Way-like galaxies, it was shown that for a particle mass between $300$ keV and $378$ keV, the fermion core can cast a shadow-like feature with a diameter of $\sim 50\ \mu$as. 

In the second part of this work, we centred on the dynamics of the G-cluster stars. These objects have relatively long orbital periods, so the data count only $\sim 2\%$ of orbital coverage. To fit the orbits of these stars, it was necessary to fix the gravitational potentials of the three models selected here. We took the best-fit values given by the maximum of the likelihood function in the S2 star case, and left free only the orbital parameters. Although for this cluster the orbital parameters and predicted quantities differ somewhat appreciably from one model to another (see Table \ref{tab:G-parameters} and figure \ref{fig:G-orbits}), it is not yet sufficient to provide conclusive evidence favouring one model over another when computing the corresponding Bayes factors (Table \ref{tab:Bayes-factors}). 

All in all, we conclude that it is necessary to have a better quality and quantity of data to differentiate between the BH and fermionic models. In addition, accurate enough data from stars orbiting inside the S2 orbit is crucial, given it tests stronger gravitational potentials in the surroundings of Sgr A*.  

\vspace{-0.5 cm}
\section*{Acknowledgements}

V.C. thanks financial support from CONICET, Argentina. 
E.A.B-V. thanks the financial support from  VIE - UIS Postdoctoral Fellowship Program No. 2025000167. 
C.R.A. is supported by the CONICET of Argentina, the ANPCyT-FONCyT (grant PICT-2022-2022-03-00332), and ICRANet.
MFM acknowledges support from Universidad Nacional de La Plata (PID G178 and G197). MFM thanks Julius Krumbiegel for help with \texttt{AlgebraOfGraphics}. 

This work used computational resources from CCAD – Universidad Nacional de Córdoba (\href{https://ccad.unc.edu.ar/}{https://ccad.unc.edu.ar/}), which are part of SNCAD – MinCyT, República Argentina.

\vspace{-0.5 cm}
\section*{Data Availability}

The astrometric data used in this work were obtained from \citet{2019Sci...365..664D,2020Natur.577..337C,2021ApJ...923...69P,2023ApJ...943..183P}. The rotation curve data were obtained from \citet{sofue_rotation_2013,Jiao2023}.

\section*{Software}
The algorithms used in this work were mainly written in the programming language \texttt{Python3}~\citep{van1995python}. We made intensive use of the
\texttt{NumPy}~\citep{harris2020array}, \texttt{SciPy}~\citep{2020SciPy-NMeth}, \texttt{emcee}~\citep{2013PASP..125..306F} and \texttt{MCEvidence}~\citep{heavens2017marginallikelihoodsmontecarlo} libraries.
Most of the figures presented in this work were made with \texttt{Matplotlib}~\citep{Hunter:2007} and the module \texttt{Corner} \citep{corner}.
For the rotation curve figures we used the Julia~\citep{bezanson2017julia} libraries: \texttt{Makie}~\citep{DanischKrumbiegel2021} and \texttt{AlgebraOfGraphics}~(\href{https://aog.makie.org/dev/}{https://aog.makie.org/dev/}).

\vspace{-0.2cm}

\bibliographystyle{mnras}
\bibliography{references} 


\vspace{-0.2cm}
\appendix
\section{Complementary figures} 

\begin{figure*}
    \centering
    \includegraphics[width=\linewidth]{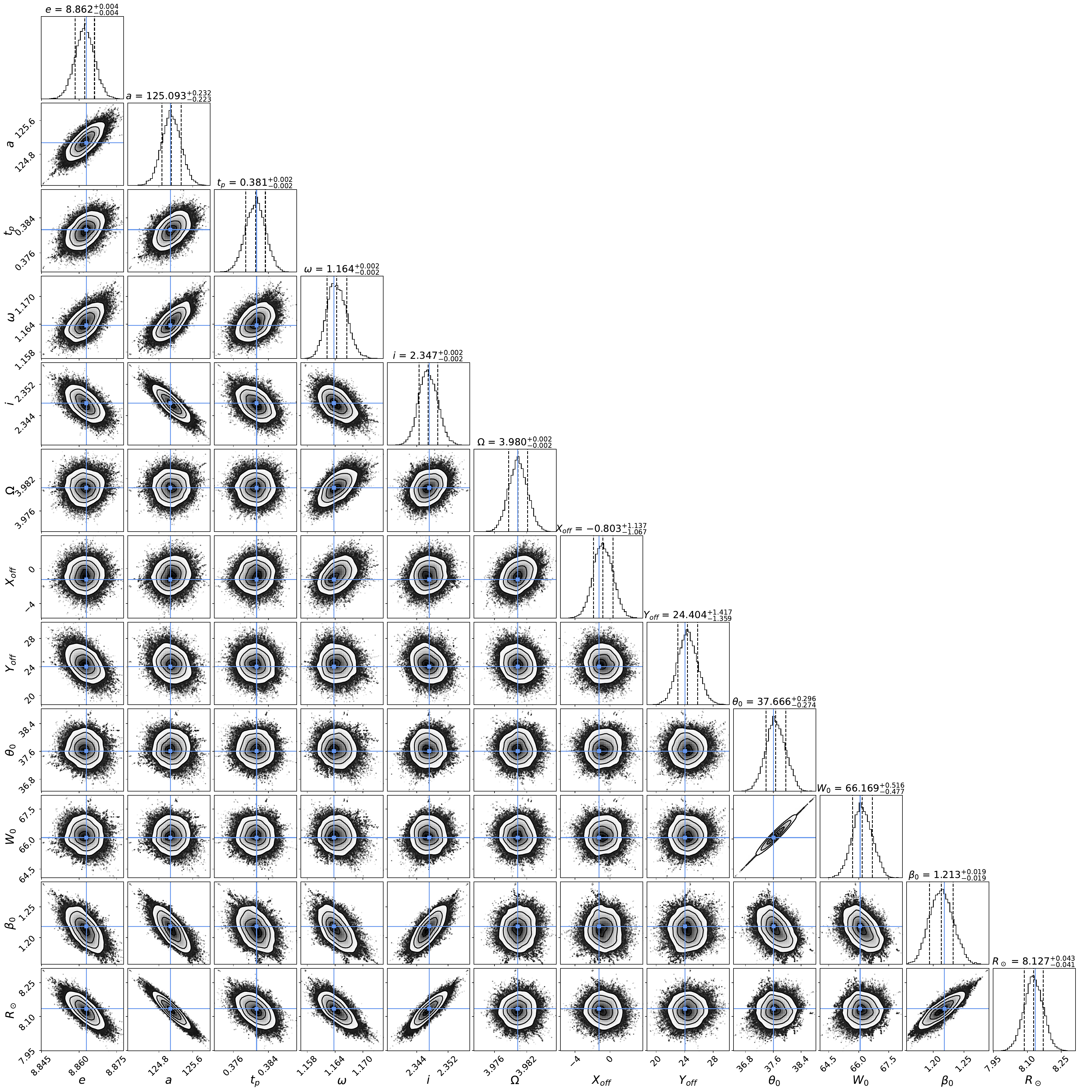}
    \caption{MCMC posteriors for the S2 star using the RAR-DM model $\mathcal{M}_{1}$ of $mc^2=56$ keV fermions. Coloured lines correspond to the best-fit values of the parameters. Dashed lines in the projected 1D posteriors correspond to the mean (reported value above each box) and $1\sigma$ significance level. For clarity, certain parameters have been rescaled. Their values are as follows $e[10^{-1}]$, $a[10^{-3}]$, $t_{\rm p} [+2018]$, $X_{\rm off}[10^{-1}]$, $Y_{\rm off}[10^{-1}]$, $\beta_0 [10^{-5}]$. }
    \label{fig:56keV cornerplot}
\end{figure*}

\begin{figure*}
    \centering
    \includegraphics[width=\linewidth]{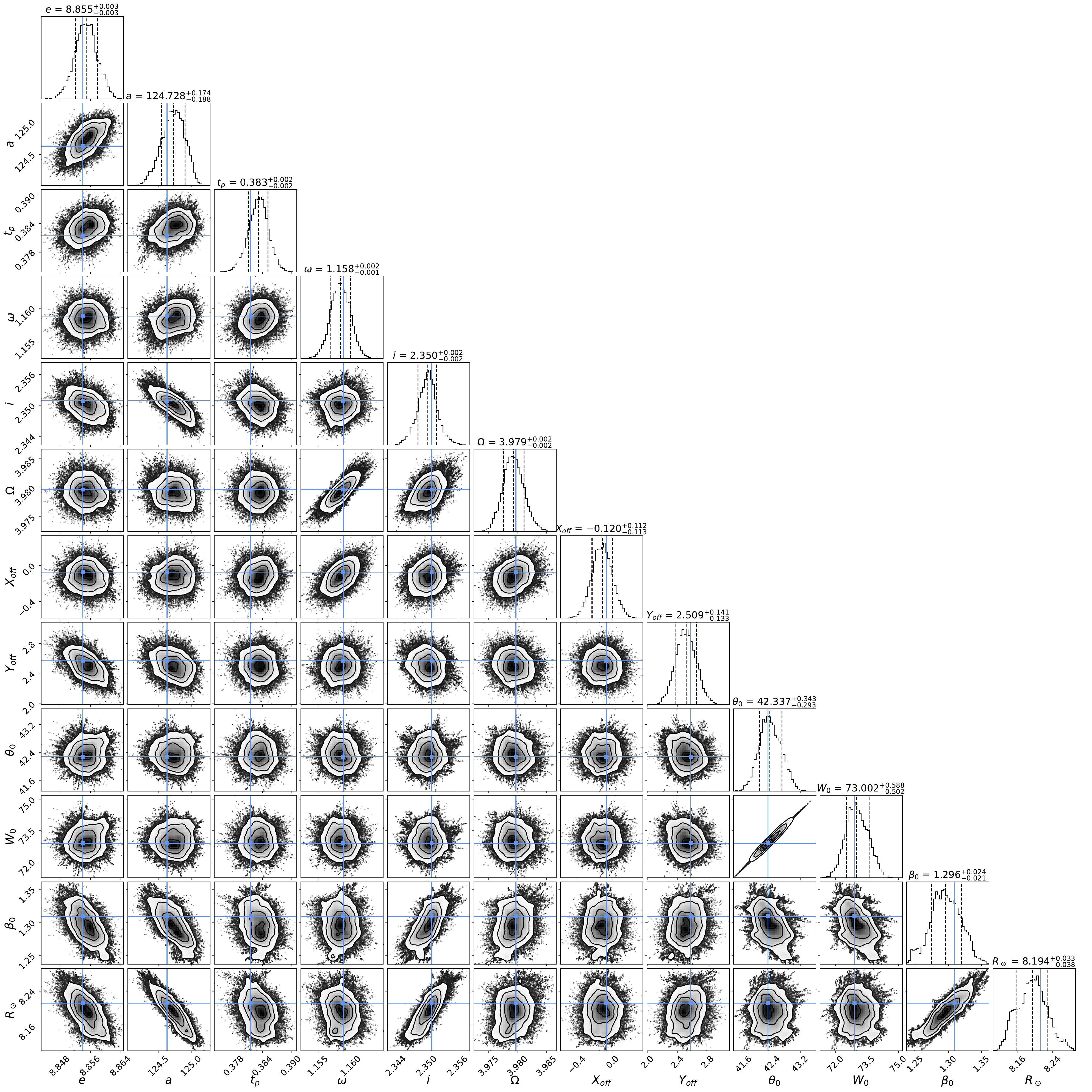}
    \caption{MCMC posteriors for the S2 star using the RAR-DM model $\mathcal{M}_{2}$ of $mc^2=300$ keV fermions. Coloured lines correspond to the best-fit values of the parameters. Dashed lines in the projected 1D posteriors correspond to the mean (reported value above each box) and $1\sigma$ significance level. For clarity, certain parameters have been rescaled. Their values are as follows $e[10^{-1}]$, $a[10^{-3}]$, $t_{\rm p} [+2018]$, $\beta_0 [10^{-3}]$.}
    \label{fig:300keV cornerplot}
\end{figure*}

\begin{figure*}
  \centering
  \includegraphics[scale=0.28]{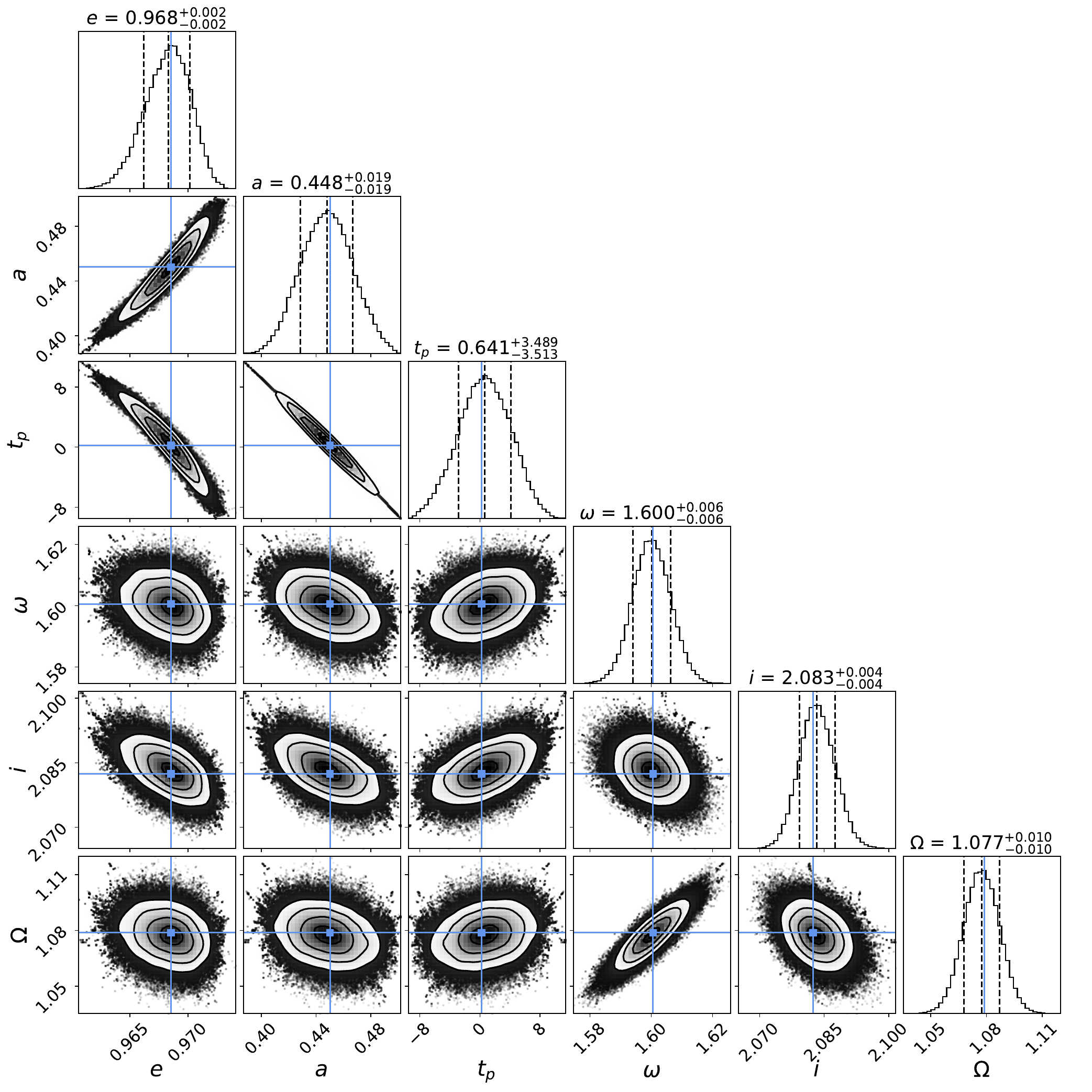}
  \caption{MCMC posteriors for the G2 star using the RAR-DM model $\mathcal{M}_{1a}$ of $mc^2=56$ keV fermions. The gravitational potential was fixed according to S2 constraints. Coloured lines correspond to the best-fit values of the parameters. Dashed lines in the projected 1D posteriors correspond to the mean (reported value above each box) and $1\sigma$ significance level. For clarity, the $t_{\rm p}$ parameter has been rescaled to $t_{\rm p} [+2014]$. }
  \label{fig:G2_MCMC}
\end{figure*}

Here we present posterior probability density corner plots. For extent reasons, we only show the full parameter space of fermionic models $\mathcal{M}_{1}$ (figure \ref{fig:56keV cornerplot}) and $\mathcal{M}_{2}$ (figure \ref{fig:300keV cornerplot}) sampled for S2 data and 2 MW halo mass-ratio constraints. Of the G-objects we only exhibit posteriors sampled for G2 (figure \ref{fig:G2_MCMC}) and G3 (figure \ref{fig:G2_MCMC}), corresponding to model $\mathcal{M}_{1a}$.
We found compact unimodal posterior distributions in all the scenarios studied in this work, i.e., 6 stars and 6 models, reflecting the robustness of the parameters.
 
\newpage

\newpage
\begin{figure*}
    \centering
    \includegraphics[width=\linewidth]{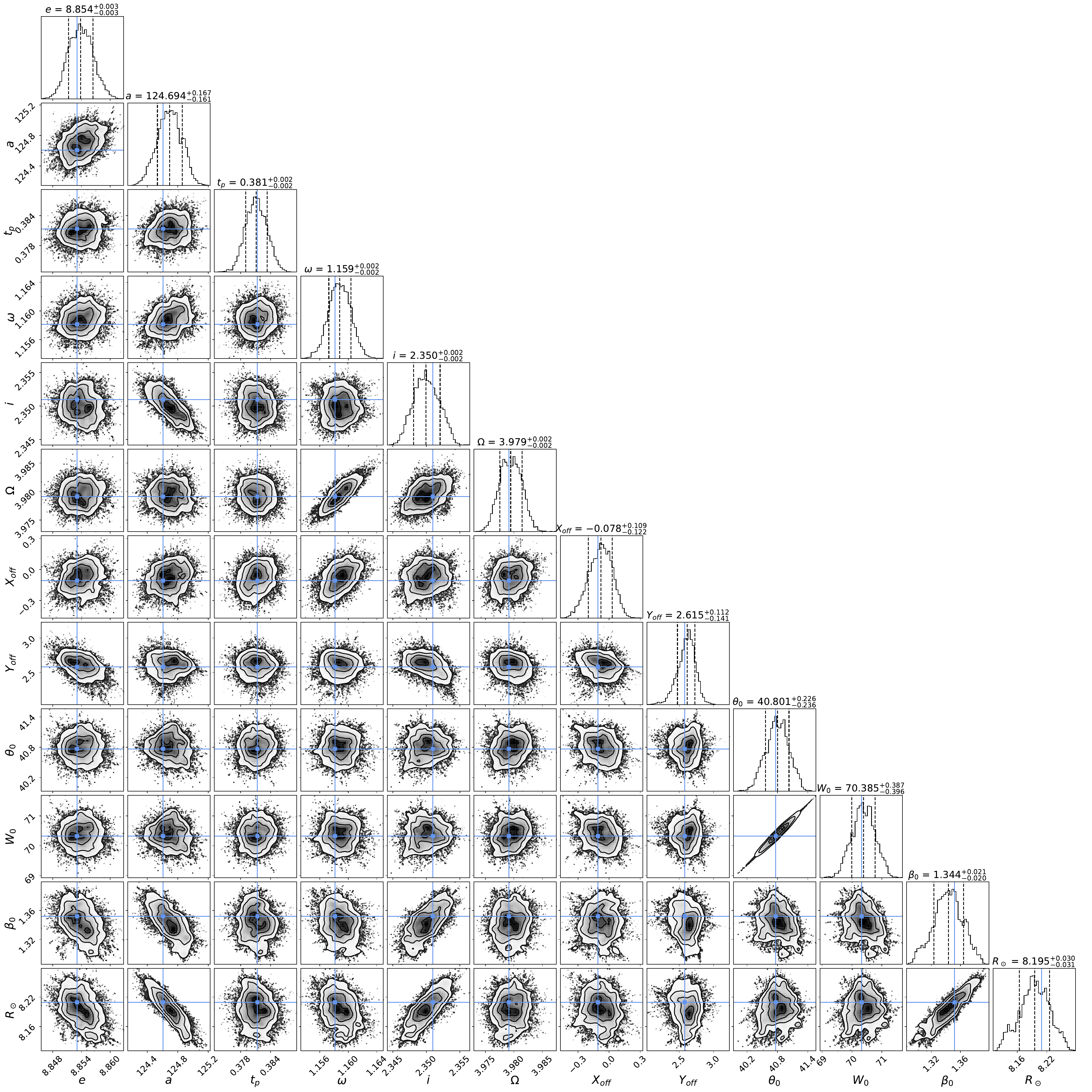}
    \caption{MCMC posteriors for the S2 star using the RAR-DM model with GAIA-DR3 halo constraints for $mc^2=300$ keV fermions. Coloured lines correspond to the best-fit values of the parameters. Dashed lines in the projected 1D posteriors correspond to the mean (reported value above each box) and $1\sigma$ significance level. For clarity, certain parameters have been rescaled. Their values are as follows $e[10^{-1}]$, $a[10^{-3}]$, $t_{\rm p} [+2018]$, $\beta_0 [10^{-3}]$. }
    \label{fig:300keV cornerplot GAIA}
\end{figure*}

\newpage
\begin{figure*}
  \centering
  \includegraphics[scale=0.28]{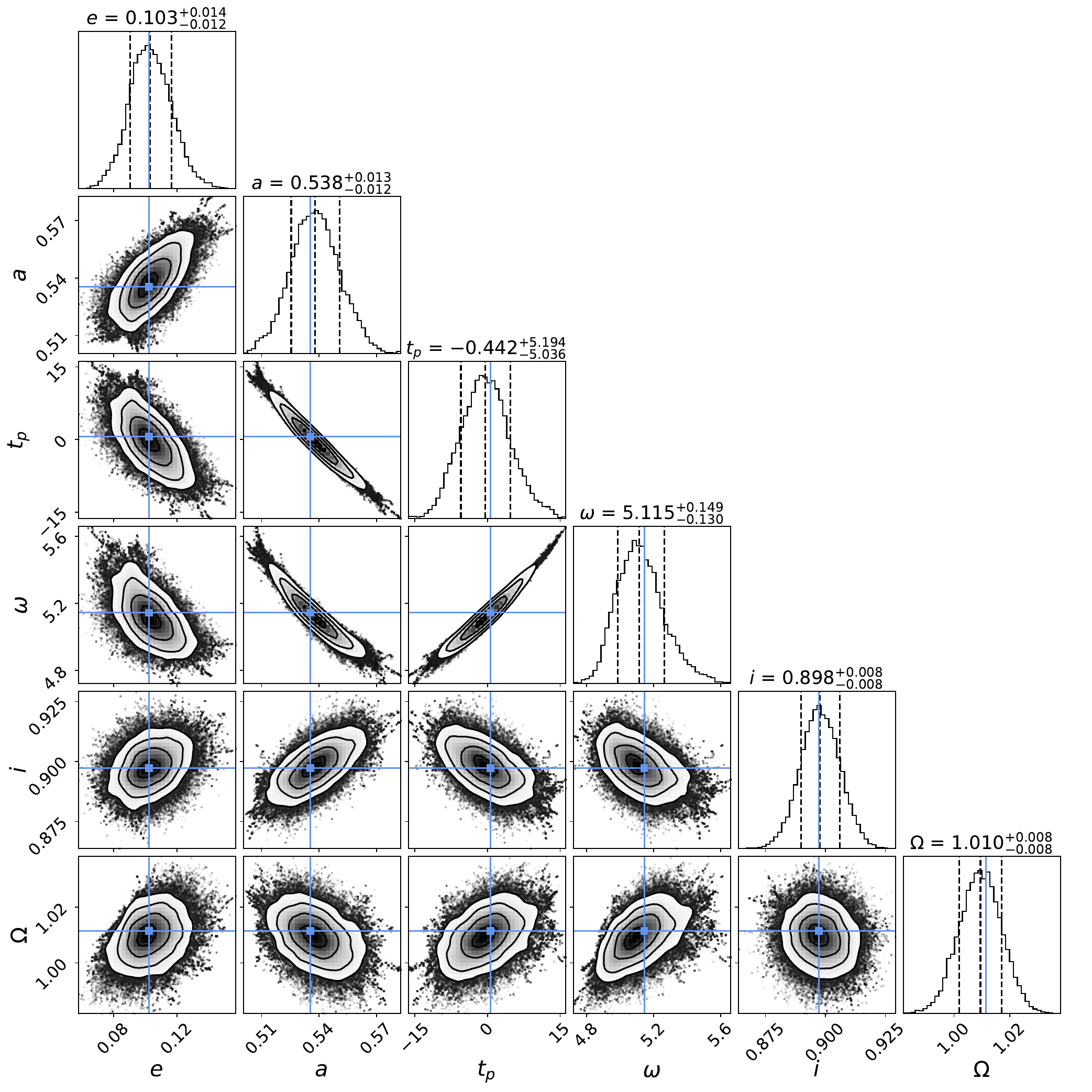}
  \caption{MCMC posteriors for the G3 star using the RAR-DM model $\mathcal{M}_{1a}$ of $mc^2=56$ keV fermions. The gravitational potential was fixed according to S2 constraints. Coloured lines correspond to the best-fit values of the parameters. Dashed lines in the projected 1D posteriors correspond to the mean (reported value above each box) and $1\sigma$ significance level. For clarity, the $t_{\rm p}$ parameter has been rescaled to $t_{\rm p} [+2035]$. }
  \label{fig:G3_MCMC}
\end{figure*}


\bsp	
\label{lastpage}
\end{document}